\documentclass{iopart}
\usepackage{iopams}
\usepackage{graphicx,amsbsy,bbm}
\usepackage{psfig}
\usepackage{multirow}
\usepackage{pstricks,pst-grad}

\newcommand{\ket}[1]{\vert #1 \rangle}

\newcommand{\mbfrac}{\mbox{$\frac12$}}

\newcommand{\bmsigma}{\boldsymbol\sigma}

\newcommand{\bmLambda}{\boldsymbol \Lambda}

\newcommand{\sT}{\scriptscriptstyle T}

\newcommand{\bmA}{{\boldsymbol A}}
\newcommand{\bmB}{{\boldsymbol B}}
\newcommand{\bmC}{{\boldsymbol C}}

\newcommand{\bmM}{{\boldsymbol M}}

\newcommand{\bmR}{{\boldsymbol R}}

\newcommand{\bmX}{{\boldsymbol X}}

\newcommand{\bmZ}{{\boldsymbol Z}}

\newcommand{\rmSU}{{\rm SU}}

\newcommand{\be}{\begin{equation}}
\newcommand{\ee}{\end{equation}}
\newcommand{\bea}{\begin{eqnarray}}
\newcommand{\eea}{\end{eqnarray}}

\newcommand{\ii}{\mathbbm{1}}

\newcommand{\jj}{\mathbb{J}}

\newcommand{\gr}[1]{\boldsymbol{#1}}

\newcommand{\refeq}[1]{Eq.~(\ref{#1})}

\renewcommand{\det}{{\rm Det}\,}

\def\re#1{\Re\hbox{e}[#1]}
\def\im#1{\Im\hbox{m}[#1]}

\newcommand{\KPsi}{|\gr{\Psi}_m\rangle}

\newcommand{\calC}{{\cal C}}
\newcommand{\bmcalN}{\boldsymbol {\cal N}}
\newcommand{\bmcalA}{\boldsymbol {\cal A}}
\newcommand{\bmcalB}{\boldsymbol {\cal B}}
\newcommand{\pp}{\mathbb{P}}
\newcommand{\GG}{\mathbb{G}}
\newcommand{\tinyT}{{\mbox{\tiny }}}

\newcommand{\on}{\overline n}
\newcommand{\oX}{\overline \bmX}
\newcommand{\Flcdt}{F_{\mbox{\tiny{\rm LCDT}}}}
\newcommand{\ba}{\begin{eqnarray}}

\begin{document}
\title{Local vs Nonlocal cloning in a noisy environment}
\author{Alessandro Ferraro and Matteo G. A. Paris}
\address{Dipartimento di Fisica dell'Universit\`a di Milano, Italy.}
\begin{abstract}
  We address the distribution of quantum information among many
  parties in the presence of noise. In particular, we consider how to
  optimally send to $m$ receivers the information encoded into an
  unknown coherent state. On one hand, a local strategy is considered,
  consisting in a local cloning process followed by direct
  transmission. On the other hand, a telecloning protocol based on
  nonlocal quantum correlations is analyzed. Both the strategies are
  optimized to minimize the detrimental effects due to losses and
  thermal noise along the propagation. The comparison between the
  local and the nonlocal protocol shows that telecloning is more
  effective than local cloning for a wide range of noise parameters.
  Our results indicate that nonlocal strategies can be more robust
  against noise than local ones, thus being suitable candidates to
  play a major role in quantum information networks.
\end{abstract}
\maketitle
\section{Introduction}\label{intro}
One of the scope of the burgeoning field of quantum information with
continuous variables (CV) \cite{vLB_rev,Napoli} is to replace, in some
particularly crucial purposes, the communication technology currently
in use. Major progresses in this field are, for instance, the
experimental realizations of quantum teleportation and cryptography
\cite{tlpqkd}.  They involve, in general, only two parties, a sender
and a receiver, and are based either on nonlocal quantum correlations,
or on protocols involving local manipulation and direct transmission
of quantum states. For example, the issue of sending an unknown state
to a single receiver has been addressed, comparing the performances of
teleportation and direct transmission \cite{mskim,ban,oli}. Of
particular interest is the case in which the channel supporting the
transfer of quantum states is affected by thermal noise and losses, as
in real experiments.  Concerning the transfer of nonclassical
features, it has been shown that entanglement is necessary in a
teleportation scenario \cite{mskim} and that teleportation is
preferable to direct transmission in certain regimes \cite{ban}.
Furthermore, a communication protocol in which information is encoded
onto the field amplitude (amplitude-modulated communication) of a set
of Gaussian pure states has been analyzed \cite{oli}, indicating that
teleportation can be more effective also in this case.  \par The
further natural step is to consider more complex communication
scenarios, where more than two parties are involved in what is called
a quantum information network. The recent experimental realizations of
CV dense coding \cite{dense} and quantum teleportation network
\cite{qtn} involve in fact three distinct parties. In this work, we
consider the problem of distributing the information encoded in an
unknown coherent state of a CV system to $m$ parties, in a
multipartite amplitude-modulated communication scenario. As for the
single receiver case, one may ask whether it is better to perform the
distribution using a local or a nonlocal strategy. On one hand, one
may in fact consider to clone the original state somewhere along the
noisy transmission line by means of an optimal local cloning machine.
In this case both the signal and the clones are directly coupled with
the environment and, then, the fidelity is affected by the unavoidable
degradation of the signal and the clones themselves.  On the other
hand, a pre-shared multipartite entangled state may be used to support
a telecloning protocol. In this case, the performance of the protocol
is affected by the degradation of the non-local correlations of the
support. The fact that optimal telecloning does not need for an
infinite amount of entanglement, as opposite to teleportation, leads
to hypothesize that the degradation of entanglement is not too
dramatic in affecting the fidelity of the clones.  \par In order to
face the effects of decoherence, both the local and the nonlocal
strategy have to be optimized. In particular, we will outline the role
of the location of the cloning machine and of the multimode state
source, both in the local and nonlocal protocol respectively.  \par
The paper is organized as follows. In Sec.~\ref{s:sum1} we will
introduce the multimode states that will be used as support of the
optimized telecloning protocol described in Sec.~\ref{s:tlc}. The
optimization of the local strategy will be outlined in
Sec.~\ref{s:clo}. Sec.~\ref{s:cfr} will be devoted to the comparison
between the two strategies, whereas the main results will be
summarized in Sec.~\ref{s:esco}.
\section{Multimode entangled states}\label{s:sum1}
Let us begin introducing the multimode entangled states that will
provide the support for the telecloning protocol. Multimode
entanglement of Gaussian states, that is states with Gaussian Wigner
function, has attracted much attention recently, both from a
theoretical and an experimental viewpoint \cite{Napoli}. A
particularly interesting class of multimode Gaussian states are the
coherent states of the group $\rmSU (m,1)$ \cite{teo_sum1,vLB_tlc}. Indeed,
this states can be experimentally generated by multimode parametric
processes in second order nonlinear crystals, with Hamiltonians that
are at most bilinear in the fields \cite{qtn,exp_sum1}. In particular,
these processes involve $m+1$ modes of the field $a_0,a_1,\dots,a_m$,
with mode $a_0$ that interacts through a parametric-amplifier-like
Hamiltonian with the other modes, whereas the latter interact one with
each other only via a beam-splitter-like Hamiltonian. The Hamiltonian
of the system is thus given by
\begin{equation}
H_m=\sum_{l<k=1}^{m} \gamma_{kl}^{(1)}\, a_k a^{\dag}_l
+ \sum_{k=1}^{m} \gamma_{k}^{(2)}\, a_k a_0 + h. c.
\label{Hm}\;,
\end{equation}
where $[a_k,a_l]=0$, $[a_k,a^\dagger_l]=\delta_{k,l}$
($k,l=0,\dots,m$) are independent bosonic modes, whereas
$\gamma_{kl}^{(1)}$ and $\gamma_{k}^{(2)}$ are coupling constants.
The states generated by $H_m$ from the vacuum are the coherent states
of the group $\rmSU (m,1)$, namely
\begin{eqnarray}
\fl
\KPsi = \sqrt{{\cal Z}_m}\sum_{\{\gr{n}\}}
\frac{\calC_1^{n_1} \calC_2^{n_2}... \calC_m^{n_m}\: \sqrt{(n_1+n_2+...
+n_m)!}}{\sqrt{n_1! n_2! ... n_m!}}\:
|\sum_{k=1}^m n_k ;\{\gr{n}\} \rangle
\label{Psi}\;,
\end{eqnarray}
where $\{\gr{n}\}=\{n_1,n_2,...,n_m\}$. The sums over $\gr{n}$ are
extended over natural numbers and, upon introducing the mean values of
the number operators $N_k=\langle a^\dag_k a_k \rangle$, we have
defined:
\begin{eqnarray}
\calC_k=\left(\frac{N_k}{1+N_0}\right)^{1/2} \;,
\qquad {\cal Z}_m=\frac{1}{1+N_0}
\qquad (k=1,\dots m)
\label{calCs}\;.
\end{eqnarray}
The relevant constant of motion, in this context, is the difference
between the mean photon number of mode $a_0$ and the total mean photon
number of the other modes. Since we start from the vacuum, we have
\begin{eqnarray}
  N_0=\sum_{k=1}^m N_k \;.
\label{cons2}
\end{eqnarray}
We notice from \refeq{Psi} that for $m=1$ the twin-beam state is
recovered. Being evolved from the vacuum with a quadratic Hamiltonian,
the states $\KPsi$ are Gaussian. They are completely characterized by
the covariance matrix $\bmsigma$, whose entries are defined as
\begin{eqnarray}
\label{defCOV}
[\bmsigma]_{kl} &=  \frac12 \langle \{R_k,R_l\} \rangle -
\langle  R_l \rangle
\langle  R_k \rangle\,,
\end{eqnarray}
where $\{A,B\}=AB+BA$ denotes the anticommutator, $\bmR=
(q_0,p_0,\ldots,q_m,p_m)^{\sT}$ and the position and momentum operator
are defined as $q_k = (a_k + a_k^\dag)/\sqrt2$ and $p_k = (a_k -
a_k^\dag)/i\sqrt2$.  The covariance matrix for the states $\KPsi$ reads
as follows:
\begin{eqnarray}
\label{CovPsi}
\bmsigma_{m} &= \left(
\begin{array}{ccccc}
\bmcalN_0 & \bmcalA_1     & \bmcalA_2     & \ldots          &  \bmcalA_m \\
\bmcalA_1 & \bmcalN_1     & \bmcalB_{1,2} & \ldots          &  \bmcalB_{1,m} \\
\bmcalA_2 & \bmcalB_{1,2} & \bmcalN_2     & \ddots          &  \vdots \\
\vdots    & \vdots        & \ddots        & \ddots          &  \bmcalB_{m-1,m} \\
\bmcalA_m & \bmcalB_{1,m} & \ldots        & \bmcalB_{m-1,m} &  \bmcalN_m \\
\end{array}
\right)\,,
\end{eqnarray}
where the entries are given by the following $2\times 2$ matrices
($k=0,\dots,m$, $h=1,\dots,m$, $j=2,\dots,m$ and $0 < i < j$)
\begin{eqnarray}
\label{CovPsiAux}
\fl
\bmcalN_k=(N_k+\frac12)\,\ii \qquad
\bmcalA_h= \sqrt{N_h(N_0+1)}\,\pp \qquad
\bmcalB_{i,j} = \sqrt{N_i\,N_j}\,\ii \;,
\end{eqnarray}
with $\ii={\rm Diag}(1,1)$ and $\pp={\rm Diag}(1,-1)$. The basic
property of the states $\KPsi$ is that they are fully inseparable,
{\em i.e.}, they are inseparable for any grouping of the modes
\cite{multitlc}. By virtue of this property, the states $\KPsi$ can
provide the support for a telecloning protocol, as we will see in the
next section.
\section{Telecloning in a noisy environment}\label{s:tlc}
\begin{figure}[h]
\begin{center}
\setlength{\unitlength}{1cm} 
\begin{picture}(15,4.3)(0,0)
\put(1.92,0.18){\includegraphics[width=10.5\unitlength]{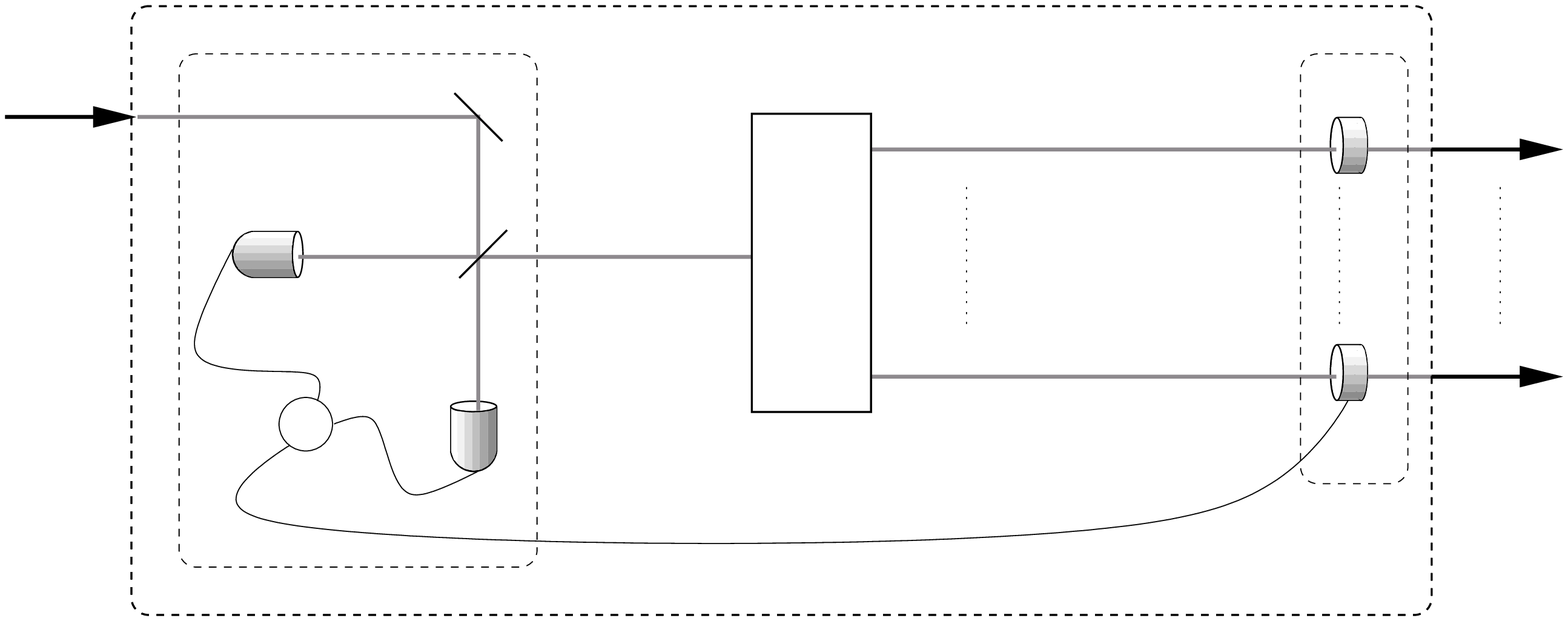}}
\put(6.1,2.6){\psline[linewidth=.04,linecolor=gray]{<-}(0,0)(0.01,0)}
\put(4.4,2.6){\psline[linewidth=.04,linecolor=gray]{<-}(0,0)(0.01,0)}
\put(4.4,3.53){\psline[linewidth=.04,linecolor=gray]{->}(0,0)(0.01,0)}
\put(5.12,3){\psline[linewidth=.04,linecolor=gray]{<-}(0,0)(0,0.01)}
\put(5.12,2){\psline[linewidth=.04,linecolor=gray]{<-}(0,0)(0,0.01)}
\put(8.8,3.32){\psline[linewidth=.04,linecolor=gray]{->}(0,0)(0.01,0)}
\put(8.8,1.8){\psline[linewidth=.04,linecolor=gray]{->}(0,0)(0.01,0)}
\put(1.9,3.8){\small{\darkgray $|\alpha\rangle $}}
\put(4,3.6){\small{\darkgray $b$}}
\put(6.96,1.56){\psframe[fillstyle=gradient,gradbegin=white,gradend=lightgray,gradmidpoint=1,gradangle=-45,linestyle=none](0,0)(.79\unitlength,1.98\unitlength)}
\put(6.95,2.5){\footnotesize{\darkgray $\KPsi$}}
\put(6.5,2.72){\small{\darkgray $a_0$}}
\put(7.9,3.45){\small{\darkgray $a_1$}}
\put(7.9,1.9){\small{\darkgray $a_m$}}
\put(3.85,1.4){\small{\darkgray $z$}}
\put(8,.3){\small{\darkgray classical channel}}
\put(10.8,.7){\small{\darkgray $U_z$}}
\put(11.6,3.45){\small{\darkgray $\varrho_1$}}
\put(11.6,1.9){\small{\darkgray $\varrho_m$}}
\put(5.5,3.7){\psline[linewidth=.01,linecolor=black]{<->}(0,0)(1.8\unitlength,0)}
\put(6,3.8){\small{\darkgray $\tau_0$}}
\put(7.3,3.7){\psline[linewidth=.01,linecolor=black]{<->}(0,0)(3.3\unitlength,0)}
\put(9,3.8){\small{\darkgray $\tau_c$}}
\put(6,2.3){\small{\darkgray $\mu$}}
\put(9,3.1){\small{\darkgray $\mu$}}
\put(9,1.5){\small{\darkgray $\mu$}}
\end{picture}
\end{center}
\caption{Schematic diagram of the telecloning scheme. After the
  preparation of the state $\KPsi$, a conditional measurement is made
  on the mode $a_0$, which corresponds to the joint measurement of the
  sum- and difference-quadratures on two modes: mode $a_0$ itself and
  another reference mode $b$, which is excited in a coherent state
  $|\alpha\rangle$, to be teleported and cloned.  The result $z$
  of the measurement is classically sent to the parties who want to
  prepare approximate clones, where suitable displacement operations
  (see text) on modes $a_1,\dots,a_m$ are performed. We indicated with
  $\mu$ the mean thermal photons in the propagation channels. The effective propagation times
  $\tau_0$ and $\tau_c$ (see text) are related to the
  losses during propagation.
\label{f:tlc}}
\end{figure}

As already mentioned, one of the main results in CV quantum
communication is the realization of the teleportation protocol. The
natural generalization of standard teleportation to many parties
corresponds to the so-called telecloning protocol \cite{murao}, {\em
  i.e.} a protocol that provides at a distance many imperfect copies
of the original input state. Teleportation is based on the coherent
states of $\rmSU(1,1)$, which provide the shared entangled states
supporting the protocol. Thus, in order to implement a multipartite
version of this protocol, one is naturally led to consider as shared
entangled state the coherent states of $\rmSU (m,1 )$ introduced in
the previous section. Actually, it has been already shown in
Ref.~\cite{multitlc} that these states permit to achieve optimal
symmetric and asymmetric telecloning of pure Gaussian states. The
telecloning protocol is schematically depicted in Fig.~\ref{f:tlc}.
After being prepared, the state $\KPsi$ propagates thorough $m+1$
noisy channels.  In particular, we can consider that modes
$a_1,\dots,a_m$ propagate in noisy channels characterized by the same
losses $\Gamma_c$. We may then define an effective propagation time
$\tau_c = \Gamma_c t$ equal for all the modes $a_1,\dots,a_m$, while
the effective propagation time $\tau_0 = \Gamma_0 t$ for mode $a_0$ is
left different from $\tau_c$. Consider in fact a scenario in which one
has two distant location (see Fig.~\ref{f:tlc}).
The distance between the two stations can be viewed as a total
effective propagation time $\tau_\tinyT$ which can be written as
$\tau_{\tinyT} =\tau_0+\tau_c$.  Then, the choice made above
corresponds to the possibility of choosing at will, for a given
$\tau_\tinyT$, which modes ($a_1,\dots,a_m$ or $a_0$) will be affected
by the unavoidable noise that separates the sending and the receiving
station and to which extent. With a slight abuse of language, we may
say that one can choose whether to put the source of the entangled
state $\KPsi$ close to the sending station ($\tau_{\tinyT}=\tau_c$),
close to the receiving one ($\tau_{\tinyT}=\tau_0$), or somewhere in
between. A similar strategy has been pursued in \cite{welsch} to
optimize the CV teleportation protocol in a noisy environment. In the
following, the optimal location and the optimal $\KPsi$ for a given
amount of noise will be given. For the sake of simplicity, all the
noisy channels will be characterized by the same effective
temperature, that is the mean thermal photons $\mu$ will be taken
equal for all the channels. As a consequence, the covariance matrix of
the evolved state is given by (see, e.g., Ref.~\cite{Napoli}):
\begin{eqnarray}
  \bmsigma_{m,{\rm n}}=\GG^{1/2}\bmsigma_m\GG^{1/2}+(1-\GG)\bmsigma_{\infty,m}\;,
\end{eqnarray}
where $\bmsigma_m$ is the initial covariance matrix of \refeq{CovPsi}
and we have defined
\begin{eqnarray}
\GG=e^{-\tau_0}\ii\oplus_{j=1}^m\,e^{-\tau_c}\ii
\qquad \bmsigma_{\infty,m}=(\mu+\mbfrac)\ii_{2m}\;.
\end{eqnarray}
Performing the calculation explicitly, upon defining
$\kappa=\mu+\frac12$, we obtain:
\begin{equation}
  \label{sigma_n}
\bmsigma_{m,\rm n}=
\left(
\begin{array}{cc}
  \bmA & \bmC \\
  \bmC\,\!^T & \bmB
\end{array}
\right)
\,,
\end{equation}
where $\bmA=e^{-\tau_0}\bmcalN_0 + \kappa(1-e^{-\tau_0})\ii$,
$\bmC = \sqrt{e^{-\tau_\tinyT}}\, (\bmcalA_1,\dots,\bmcalA_m)$ and
\begin{eqnarray}
\hspace{-1.5cm}
\fl
\bmB =
\left(
\begin{array}{ccccc}
e^{-\tau_c}\bmcalN_1 + \kappa(1-e^{-\tau_c})\ii     & e^{-\tau_c}\bmcalB_{1,2} & \ldots          & e^{-\tau_c} \bmcalB_{1,m} \\
e^{-\tau_c}\bmcalB_{1,2} & e^{-\tau_c}\bmcalN_2  + \kappa(1-e^{-\tau_c})\ii   & \ddots          &  \vdots \\
\vdots        & \ddots        & \ddots          &  e^{-\tau_c}\bmcalB_{m-1,m} \\
e^{-\tau_c}\bmcalB_{1,m} & \ldots        &e^{-\tau_c} \bmcalB_{m-1,m} &  e^{-\tau_c}\bmcalN_m + \kappa(1-e^{-\tau_c})\ii\\
\end{array}
\right)\,.\nonumber
\end{eqnarray}
As in the case of standard teleportation, the telecloning protocol now
proceeds by performing a joint measurement on modes $a_0$ and $b$,
which is excited in the unknown coherent state $|\alpha\rangle$ that we want
to teleport and clone. The measurement corresponds to an ideal
double-homodyne detection of the complex photocurrent $Z = b +
a_0^\dag$, described by the following Gaussian characteristic
function:
\begin{eqnarray}
  \chi[\bmM,\bmX](\bmLambda)=
  \exp\left\{-\frac12\bmLambda^T\bmM\bmLambda-i\bmLambda^T\bmX\right\}\;.
\label{chihom}
\end{eqnarray}
In \refeq{chihom} the covariance matrix $\bmM$ and the
vector of first moments $\bmX$ are given by:
\begin{eqnarray}
  \bmM=\pp\bmsigma_{\rm in}\pp \,,\qquad \bmX=\pp\overline\bmX+\bmZ
  \,,
\end{eqnarray}
where $\bmZ=\{\re z,\im z\}$ is the measurement result, $\bmsigma_{\rm
  in}=\frac12\ii$ and $\overline\bmX=\{\re \alpha,\im \alpha\}$ are
  the covariance matrix and the vector of first moments of the input
  coherent state $\ket{\alpha}$. Then, the state $\varrho_c$,
  conditioned to the result $\bmZ$, is a Gaussian state with
  covariance matrix
\begin{eqnarray}
  \bmsigma_c=\bmB-\bmC^T(\bmA+\bmM)^{-1}\bmC
\end{eqnarray}
and vector of first moments ${\boldsymbol
  H}=\bmC^T(\bmA+\bmM)^{-1}\bmX$. 
The probability $P(\bmZ)$ of the outcome $\bmZ$ is given by:
\begin{eqnarray}
  P(\bmZ)=\frac{1}{\sqrt{\det(\bmA+\bmM)}}\exp\left\{
-\frac12 \bmX^T(\bmA+\bmM)^{-1}\bmX
\right\}\,.
\end{eqnarray}
After the measurement, the conditional state should be transformed by
a further unitary operation, depending on the outcome of the
measurement. In our case, this is a $m$-mode product displacement $U_z
= \bigotimes_{h=1}^m D_h^{\sT}(z)$. This is a local transformation,
which generalizes to $m$ modes the procedure already used in the
original CV teleportation protocol. The characteristic function of the
modes $a_1,\dots,a_m$ is now given by:
\begin{eqnarray}
\chi[U_z\,\varrho_c\,U^\dagger_z](\bmLambda)=
\chi[\varrho_c](\bmLambda)\exp\left\{i\bmLambda^T\jj^T\bmZ^* \right\} \;,
\end{eqnarray}
where $\bmLambda$ is the usual $2m$-component column vector spanning
the reciprocal phase space of modes $a_1,\dots,a_m$, whereas $^*$
indicates complex conjugation and $\jj$ is given by the $2\times2m$
matrix $\jj=(\ii,\dots,\ii)$. The characteristic function of the
overall output state $\varrho_{out}$ is obtained by averaging over all
the possible outcomes
\begin{eqnarray}
\fl
  \chi_{out}(\bmLambda)&
\!\!\!\!\!\!\!\!\!\!\!\!\!\!\!\!\!\!\!\!\!\!\!
=\int d^{2m}\bmZ \,P(\bmZ)\,
\chi[\varrho_c](\bmLambda)\,\exp\!\left\{i\bmLambda^T\jj^T\bmZ^* \right\} \\
&
\!\!\!\!\!\!\!\!\!\!\!\!\!\!\!\!\!\!\!\!\!\!\!
=\exp\left\{-\mbfrac\bmLambda^T
\left[
\bmB+\jj^T\pp(\bmA+\bmM)\pp\jj-\jj^T\pp\bmC-\bmC^T\pp\jj
\right]\bmLambda
-i\bmLambda^T\jj^T\overline\bmX
\right\}\,,
\end{eqnarray}
which, in turn, gives the following covariance matrix for the $h$-th
clone $\varrho_{h}={\rm Tr}_{i\neq h}[\varrho_{out}]$:
\begin{eqnarray}
\label{Reduced_n}
\bmsigma_h=
\left(\frac{1}{F_h}-\frac12\right)
\ii\,.
\end{eqnarray}
In the Equation above, $F_h$ represents the fidelity
$F_h=\langle\alpha|\varrho_h|\alpha\rangle$ between the $h$-th clone
and the original coherent state, {\em i.e.}:
\begin{eqnarray}
\label{FidNoise}
\fl
F_h &
\!\!\!\!\!\!\!\!\!\!\!\!\!\!\!\!\!\!\!\!\!\!\!\!\!\!\!\!\!\!\!\!\!\!\!
=\left\{\det\left[\bmsigma_h+\mbfrac\ii
  \right]\right\}^{-1/2} \nonumber \\
&\!\!\!\!\!\!\!\!\!\!\!\!\!\!\!\!\!\!\!\!\!\!\!\!\!\!\!\!\!\!\!\!\!\!\!
= \left\{2+2\mu+\left[ e^{-\tau_0}\left(N_0-\mu\right)
    +e^{-\tau_\tinyT+\tau_0}\left(N_h-\mu\right)
    -2\,\sqrt{e^{-\tau_\tinyT}\,N_h(N_0+1)}\right]
\right\}^{-1} \!\!\,,
\end{eqnarray}
where we have reintroduced the mean thermal photons $\mu$. Notice that
the fidelity does not depend on the amplitude $\alpha$ of the input
state. Remarkably, from \refeq{FidNoise} follows that the present
telecloning scheme is able to perform both symmetric and asymmetric
distribution of information \cite{multitlc}. However, being interested
in the comparison between the performances of telecloning and of the
local strategy that will be outlined in the next section, from now on
we will consider only the symmetric instance. Namely, we set
$N_1=\dots=N_m=N$ and $N_0=m\,N$, from which it follows that all the
clones are equal one to each other ($F_1=\dots=F_m=F$).  \par The next
step is now to optimize, for a fixed amount of noise, the shared state
$\KPsi$ and the location of its source between the sending and the
receiving station. Namely, relying upon the fidelity as the relevant
figure of merit, one has to find the optimal $N$ and $\tau_0$ which
maximize $F$ for $\tau_\tinyT$ and $\mu$ fixed. The result of the
optimization are summarized in Tab.  \ref{optim} \cite{multitlc},
where we have defined the following quantities:
\begin{eqnarray}
F^a &= \frac{m}{
m\left[2+\mu(1-e^{-\tau_\tinyT})
\right]-1} \label{Fa}\,,\\
F^b &= \left[2+\mu-(1+\mu)e^{-\tau_\tinyT}
\right]^{-1} \label{Fb}\,,\\
F^c &= \bigg\{
2+2\mu-\sqrt{e^{-\tau_\tinyT}/m}\left[1+\mu(1+m)
\right]   \bigg\}^{-1} \label{Fc}\,.
\end{eqnarray}
\begin{table}[h]
\begin{center}
\begin{tabular}{|c|c|c|c|c|}
\hline $\mu$ & $\tau_\tinyT$  & $\tau_0^{\rm opt}$ & $N^{\rm opt}$ &
$F^{\rm max}$ \vspace{0cm}\\
\hline\hline
\rule[-4mm]{0mm}{1.1cm} 
$\forall \mu$ & 
$0<\tau_\tinyT<\ln m$ & 
$\tau_\tinyT$ & 
${\displaystyle \frac{1}{m(m\,e^{-\tau_\tinyT}-1)}}$ & 
$F^a$ \\
\hline
\multirow{2}*{\rule[-2mm]{0mm}{1cm} $\mu<\frac{1}{m-1}$} 
\rule[-4mm]{0mm}{1.1cm}& 
${\displaystyle \ln m<\tau_\tinyT<\ln\left[\frac{(1+\mu)^2}{m\,\mu^2}\right]}$ & 
$\mbfrac(\tau_\tinyT+\ln m)$ & 
$N\rightarrow\infty$ &
$F^c$ \\
\cline{2-5} & \rule[-4mm]{0mm}{1.1cm}
$\tau_\tinyT>\ln \frac{(1+\mu)^2}{m\,\mu^2}$ & 
$\tau_\tinyT$ & 
${\displaystyle \frac{e^{-\tau_\tinyT}}{1-m\,e^{-\tau_\tinyT}}
}$ &
$F^b$ \\
\hline
\rule[-4mm]{0mm}{1.1cm}
$\mu>\frac{1}{m-1}$ & 
$\tau_\tinyT>\ln m$ & 
$\tau_\tinyT$ & 
${\displaystyle \frac{e^{-\tau_\tinyT}}{1-m\,e^{-\tau_\tinyT}}}$ &
$F^b$ \\
\hline
\end{tabular}
\end{center}
\caption{ Values of the optimized $N^{\rm opt}$ and $\tau_0^{\rm
opt}$ for fixed values of $\tau_\tinyT$ and $\mu$. The value reached
by the fidelity $F^{\rm max}$ for these optimal choices is given
in the last column. \label{optim}}
\end{table}
\par
The most interesting feature which emerges from an inspection of
Tab.~\ref{optim} is that telecloning saturates the bound for optimal
cloning \cite{cerf} even in the presence of losses, for propagation
times $\tau_\tinyT<\ln m$, hence divergent as the number of modes
increases.  More specifically, consider the first row in
Tab.~\ref{optim} and set $\mu=0$. Then, one has that for
$\tau_\tinyT<\ln m$ the maximum fidelity is given by $F^{\rm
  max}=m/(2m-1)$. That is, the optimal fidelity for a symmetric
cloning can still be attained, carefully choosing $N$ and $\tau_0$.
This is due to the fact that multimode entanglement is robust against
this type of noise and, even if decreased along the transmission line,
it is still sufficient to provide optimal cloning. Actually, as we
already mentioned, there is no need of an infinite amount of
entanglement to perform an optimal telecloning process \cite{vLB_tlc}. Notice
that when $F^{\rm max}=F_b$ in Tab.~\ref{optim}, telecloning ceases to
be interesting, because $F_b$ is lower than the so called classical
limit $F=\frac12$ \cite{benchmark}. It is in fact immediate to see
that $F_b>\frac12$ only when
\begin{eqnarray}
\label{UnFb}
  \tau_\tinyT<\ln\left(1+\frac1\mu\right)\,.
\end{eqnarray}
However, from the third row of Tab.~\ref{optim} one has that
$\tau_\tinyT>\ln \frac{(1+\mu)^2}{m\,\mu^2}$ and $\mu<\frac{1}{m-1}$,
which implies that Ineq.~(\ref{UnFb}) cannot be satisfied. The same
conclusion holds also if one consider the forth row of
Tab.~\ref{optim}. The comparison of the results in Tab.~\ref{optim}
with the performances of a local distribution of information will be
given in Sec.~\ref{s:cfr}.
\section{Local cloning plus direct transmission (LCDT)}\label{s:clo}
Let us now consider the situation in which the distribution of quantum
information does not rely upon sharing any entanglement between the
parties involved. We refer to this kind of protocols as local cloning
+ direct transmission (LCDT) schemes. In the notation introduced in
Sec.~\ref{s:tlc}, one has that the sending and the receiving stations
are separated by an effective time $\tau_\tinyT$. The input coherent
state, characterized by the covariance matrix
$\bmsigma_{in}=\frac12\ii$ and the amplitude $\overline X$, propagates
through a noisy channel for time $\tau_0$, after which it is cloned by
a suitably chosen local optimal symmetric cloning machine. Then, the
clones are sent to the receiving station, via $m$ noisy channels for a
propagation time $\tau_c$. Before calculating the fidelity of such
LCDT strategy, it is necessary to identify the proper cloning machine
to be used. The natural requirement for a coherent state cloning
machine is its covariance with respect to displacements in the phase
space \cite{cerf23}.  This implies that the cloning map is a Gaussian noise
map of the form:
\begin{eqnarray}
  \label{gnoise}
\varrho_{\rm clo}=\frac{1}{\pi \on}\int d^2\beta\, e^{-|\beta|^2/\on}\,D(\beta)\varrho_{\rm in}D^\dagger(\beta)\;,
\end{eqnarray}
being $\varrho_{\rm in}$ the density matrix of the state at the
input of the cloning machine and $\on$ is the noise added by the
cloning process. The density matrix of the clones $\varrho_{\rm clo}$
is the partial trace over all the modes except one of the overall
state $R$ at the output of the cloning machine, namely   $\varrho_{\rm
clo}={\rm Tr}_{a_2,\dots,a_m}[R]$ (recall that we are considering the
case in which the clones are all equal). Actually, the overall state
$R$ plays no role in our analysis. Indeed, once the partial traces
$\varrho_{\rm clo}$ are fixed by the requirement in \refeq{gnoise},
the overall state $R$ has no influence on the clones propagating
through the noisy channels. In fact, the $m$ noisy channels are
independent, and the overall Liouvillian superoperator $\cal L$
factorizes into the single-channel superoperators ${\cal L}_h$. As a
consequence, one can easily show, considering the Kraus decomposition
of each ${\cal L}_h$, that:
\begin{eqnarray}
  \varrho_l={\rm Tr_{a_2,\dots,a_m}}[{\cal L}(R)]={{\cal
      L}_1}({\rm Tr_{a_2,\dots,a_m}}[R])={\cal
      L}_1(\varrho_{\rm clo})\;,
\end{eqnarray}
where $\varrho_l$ is the final state of the clones. The remaining step
to be performed is now the optimization of the location of the cloning
machine. To this aim, let us calculate the fidelity between the clones
at the end of the transmission line and the input state. After
propagating for a time $\tau_0$ the input state covariance matrix and
amplitude are given by
\begin{eqnarray}
  \bmsigma^{\rm in}_{\rm clo}=[\mbfrac+(1-e^{-\tau_0})\mu]\ii\;, \qquad
\bmX^{\rm in}_{\rm clo}=e^{-\tau_0/2}\oX\;.
\end{eqnarray}
Then, the cloning machine produces $m$ optimal clones ($\on=(m-1)/m$)
accordingly to \refeq{gnoise}, {\em i.e.}
\begin{eqnarray}
  \bmsigma^{\rm out}_{\rm clo}=[\mbfrac+(1-e^{-\tau_0})\mu+\on]\ii\;,
  \qquad \bmX^{\rm out}_{\rm clo}=e^{-\tau_0/2}\oX\;.
\end{eqnarray}
Letting the latter propagate one finally has
\begin{eqnarray}
  \bmsigma_l=[\mbfrac+(1-e^{-\tau_\tinyT})\mu+\on\,e^{-\tau_c}]\ii\;,
  \qquad \bmX_l=e^{-\tau_\tinyT/2}\oX\;,
\label{cloprop}
\end{eqnarray}
from which the fidelity $F_d$ follows:
\begin{eqnarray}\fl
  F_d=\frac{1}{1+\on\,e^{-\tau_c}+(1-e^{-\tau_\tinyT})\mu}
  \exp\left\{-\frac12\,\frac{(1-e^{-\tau_\tinyT/2})^2|\alpha|^2}{1+\on\,e^{-\tau_c}+(1-e^{-\tau_\tinyT})\mu}
  \right\}\;.
\label{Fl}
\end{eqnarray}
The maximum of $F_d$ is given by
\begin{eqnarray}
  F_d^{\rm max}=\frac{2\,e^{\tau_\tinyT-1}}{|\alpha|^2(e^{\tau_\tinyT/2}-1)^2}
\label{Flmax}
\end{eqnarray}
and it is attained for
\begin{eqnarray}
  \tau_c^{\rm opt}=\tau_\tinyT-\ln\left\{\frac{1}{2\,\on}
\left[
|\alpha|^2(e^{\tau_\tinyT/2}-1)^2+2\,\mu-2\,e^{\tau_\tinyT}(1+\mu)
\right]
\right\}\;.
\label{tcopt}
\end{eqnarray}
However, from \refeq{tcopt} it follows that $\tau_c^{\rm opt}$ is admissible (namely, $\tau_c^{\rm opt}<\tau_\tinyT$) only when
\begin{eqnarray}
  |\alpha|^2>|\tilde\alpha|^2=
  \frac{2[\on-\mu+e^{\tau_\tinyT}(1+\mu)]}{(e^{\tau_\tinyT/2}-1)^2}\,.
\end{eqnarray}
The equation above, in turn, implies that $F_d^{\rm max}$ is always lower
than the classical bound $F=\frac12$. In fact, one has that
\begin{eqnarray}
  F_d^{\rm
    max} \le\frac{2\,e^{\tau_\tinyT-1}}{|\tilde\alpha|^2(e^{\tau_\tinyT/2}-1)^2}
   =\frac1e\,\frac{e^{\tau_\tinyT}}{e^{\tau_\tinyT}+[\on+\mu(e^{\tau_\tinyT}-1)]}
  <\frac1e\,.
\end{eqnarray}
In other words, we have that, when the LCDT strategy is useful, the
best location of the cloning machine is at the sending station
($\tau_c=\tau_\tinyT$). This is due to the fact that the noisy
propagation after the cloning machine, besides degrading the signal,
decreases the noise added by the cloning machine, as we can see from
the term $\on\,e^{-\tau_c}$ in \refeq{cloprop}. The fidelity $F_d$ of
the clones produced by the optimal local strategy is thus given by:
\begin{eqnarray}\fl
 F_d=\frac{m}{m[1-\mu+e^{\tau_\tinyT}(1+\mu)]-1}
  \exp\left\{\tau_\tinyT-\frac12\,\frac{m(1-e^{\tau_\tinyT/2})^2|\alpha|^2}{m[1-\mu+e^{\tau_\tinyT}(1+\mu)]-1}
  \right\}\;,
\end{eqnarray}
which shows that the fidelity depends on the original input state. In
a communication scenario, in which the information is
amplitude-modulated, it is thus necessary to introduce a fidelity
averaged over all the possible input. Let us suppose that the message
we want to transmit is encoded in an alphabet distributed accordingly
to a Gaussian probability density function of variance $\Omega^2$:
\begin{eqnarray}
  G_\Omega(\alpha)=\frac{1}{\pi\,\Omega^2}\,e^{-|\alpha|^2/\Omega^2}\,.
\end{eqnarray}
The averaged fidelity $\Flcdt=\int d^2\alpha
\,G_\Omega(\alpha)F_d$ of the clones is thus given by:
\begin{eqnarray}
  \label{Floc}
\Flcdt =\frac{m\,e^{\tau_\tinyT}}{m[1-\mu+\Omega^2(1-2e^{\tau_\tinyT/2})+e^{\tau_\tinyT}(1+\mu+\Omega^2)]}\,.
\end{eqnarray}
The result above will be compared with the telecloning fidelity in the
next section.
\section{Comparison between local and nonlocal strategy}\label{s:cfr}
\begin{figure}[h]
\begin{center}
\setlength{\unitlength}{1cm} 
\begin{picture}(15,4.3)
\put(1.92,0.18){\includegraphics[width=5\unitlength]{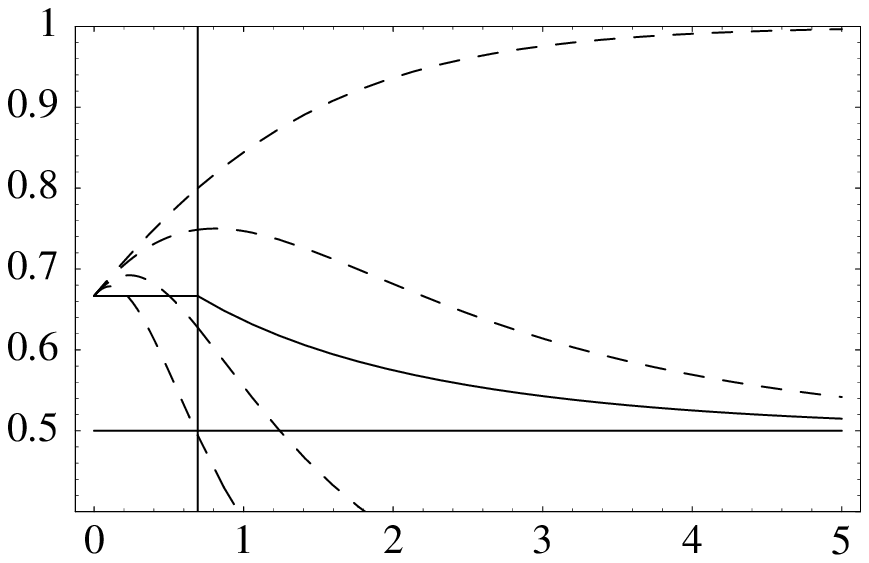}}
\put(1.8,3){\small{$F$}}
\put(6.2,0){\small{$\tau_\tinyT$}}
\put(6.2,2.7){\small(a)}
\put(7.92,0.18){\includegraphics[width=5\unitlength]{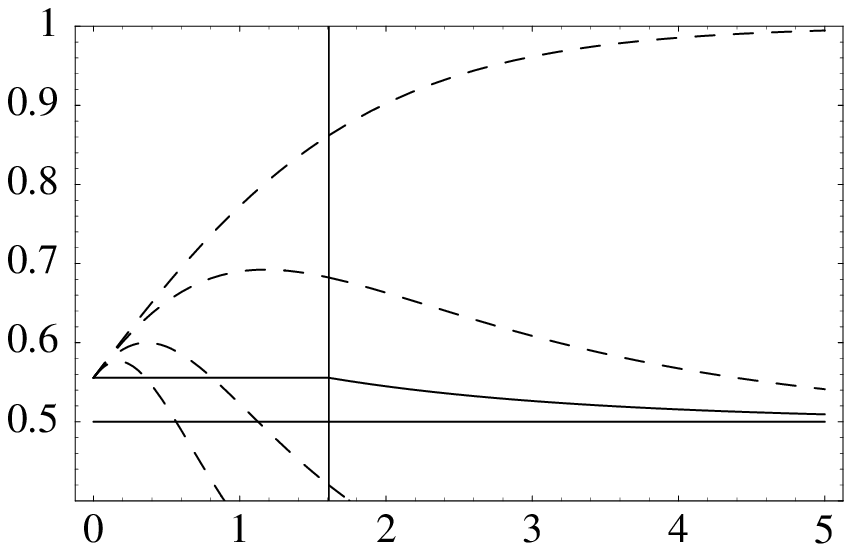}}
\put(7.8,3){\small{$F$}}
\put(12.2,0){\small{$\tau_\tinyT$}}
\put(12.2,2.7){\small(b)}
\end{picture}
\end{center}
\caption{ 
  Comparison for $\mu=0$ and $m=2$ (a) [$m=5$ (b)] between the telecloning fidelity
  given in Eq.~(\ref{FacLosses}) (solid line) and the fidelity of the
  LCDT strategy given in \refeq{Floc} (dotted lines). The latter
  refer to the case of $\Omega=0,1,2,3$ from top to bottom. The
  vertical line corresponds to $\tau_\tinyT=\ln 2$ (a) [$\tau_\tinyT=\ln 5$ (b)].
\label{f:mu0}}
\end{figure}

\begin{figure}[h]
\begin{center}
\setlength{\unitlength}{1cm} 
\begin{picture}(15,4.3)
\put(1.92,0.18){\includegraphics[width=5\unitlength]{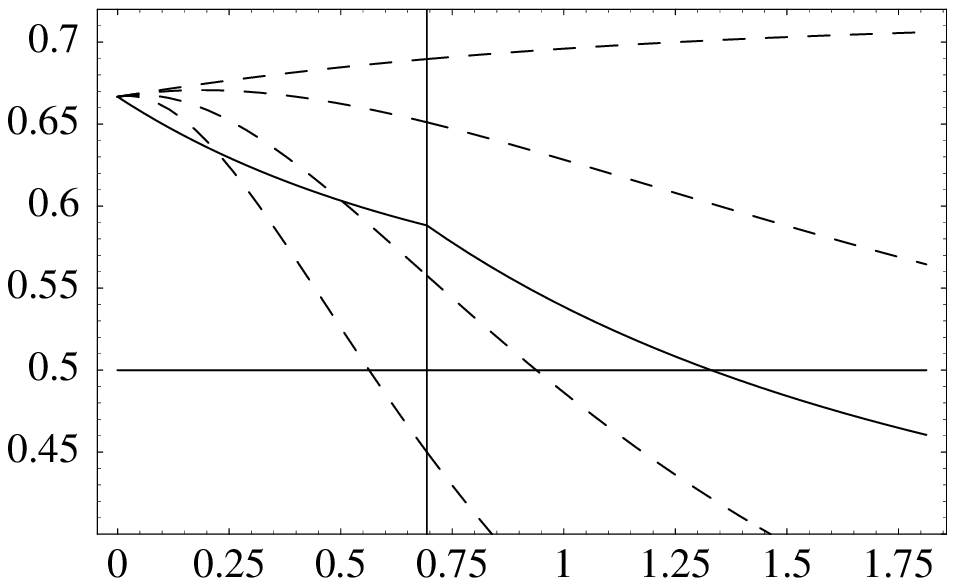}}
\put(1.8,3){\small{$F$}}
\put(6.2,0){\small{$\tau_\tinyT$}}
\put(6.2,2.7){\small(a)}
\put(7.92,0.18){\includegraphics[width=5\unitlength]{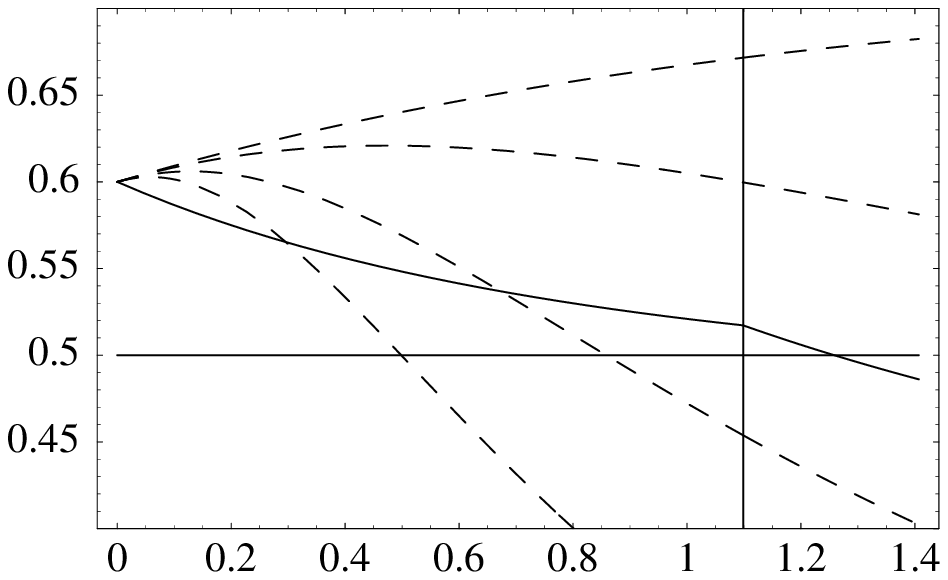}}
\put(7.8,3){\small{$F$}}
\put(12.2,0){\small{$\tau_\tinyT$}}
\put(12.2,2.7){\small(b)}
\end{picture}
\end{center}
\caption{ 
  Comparison for $\mu=0.4$ and $m=2$ (a) [$m=3$ (b)] between the telecloning fidelity
  given in Eqs.~(\ref{Fa},\ref{Fc}) (solid line) and the fidelity of the
  LCDT strategy given in \refeq{Floc} (dotted lines). The latter
  refer to the case of $\Omega=0,1,2,3$ from top to bottom. The
  vertical line corresponds to $\tau_\tinyT=\ln 2$ (a) [$\tau_\tinyT=\ln 3$ (b)].
\label{f:mu04}}
\end{figure}

We are now in the position to compare the performances of LCDT
strategy and telecloning. First, notice that the fidelity $\Flcdt$ in
\refeq{Floc} goes to zero as $\Omega$ increases.  This means that for
a truly random distributed coherent state the local strategy is not
useful at all. On the other hand, Eqs.~(\ref{Fa}-\ref{Fc}) explicitly
show that the performances of telecloning do not depend on the value
of the coherent amplitude, as it follows from the covariance of the
process. Hence, as one may expect, telecloning is undoubtedly more
effective than the LCDT strategy in case of a generic unknown coherent
state. Let us now consider the case of finite $\Omega$ in the absence
of thermal noise ($\mu=0$). Then the fidelity is given by $F_a$ and
$F_c$ in Eqs.~(\ref{Fa},\ref{Fc}), which specialize as follows:
\begin{eqnarray}
  F_a=\frac{m}{2\,m-1}\,, \qquad
  F_c=\frac{1}{2-\sqrt{e^{-\tau_\tinyT}/m}} \,.
\label{FacLosses}
\end{eqnarray}
\refeq{FacLosses} implies, as already pointed out, that optimality is still
achieved for a time $\tau_\tinyT<\ln m$, and also that the fidelity is
greater than the classical bound at any time. The comparison between
\refeq{FacLosses} and \refeq{Floc} is given in Fig.~\ref{f:mu0} for $m=2,5$. 
The figure shows that, even for small values of the width
$\Omega\simeq 2$, telecloning is more effective than the LCDT
strategy. A similar behavior is found also when thermal noise is
considered ($\mu\neq0$), as Fig.~\ref{f:mu04} shows. Notice
that in the latter case, as one may expect, also telecloning may not
give a better fidelity than the classical limit. This happens for
propagation time $\tau_\tinyT$ larger than the threshold
\begin{eqnarray}
  \tau_\tinyT^{a,{\rm th}}=\ln\left[
\frac{(1+\mu+m\,\mu)^2}{4\,m\,\mu^2}
\right]
\end{eqnarray}
for $\mu<\frac{1}{m-1}$, and larger than
\begin{eqnarray}
  \tau_\tinyT^{c,{\rm th}}=-\ln\left[
1-\frac{1}{m\,\mu}
\right]
\end{eqnarray}
otherwise.
\par
From a quantum communication viewpoint it is interesting to consider
the threshold for $\Omega$ above which telecloning becomes more effective than
LCDT strategy. The latter can be analytically retrieved and one has that
$F_a$ in \refeq{Fa} is greater than $\Flcdt$ when
$\Omega>\Omega_{a,{\rm th}}$, with
\begin{eqnarray}
  \Omega_{a,{\rm th}}^2=\frac{(1+e^{\tau_\tinyT/2})(m-1)}{(e^{\tau_\tinyT/2-1})m}\,,
\label{th_a}
\end{eqnarray}
whereas $F_c$ in \refeq{Fc} is greater than $\Flcdt$ when $\Omega>\Omega_{c,{\rm th}}$, with
\begin{eqnarray}\fl
 \Omega_{c,{\rm th}}^2=
\frac{1 + m\,\left(\mu-1\right)  + 
    m\,e^{\tau_\tinyT}\,\left( 1 + \mu \right)  - 
    \sqrt{m}\,e^{\tau_\tinyT/2}\,
     \left( 1 + \mu + m\,\mu \right) }{{m\,\left( 
        e^{\tau_\tinyT/2}-1\right) }^2}\,.
\label{th_c}
\end{eqnarray}
Notice that $\Omega_{a,{\rm th}}$ does not depend on the thermal
photons $\mu$. 
\begin{figure}[h]
\begin{center}
\setlength{\unitlength}{1cm} 
\begin{picture}(15,4.3)
\put(1.92,0.18){\includegraphics[width=5\unitlength]{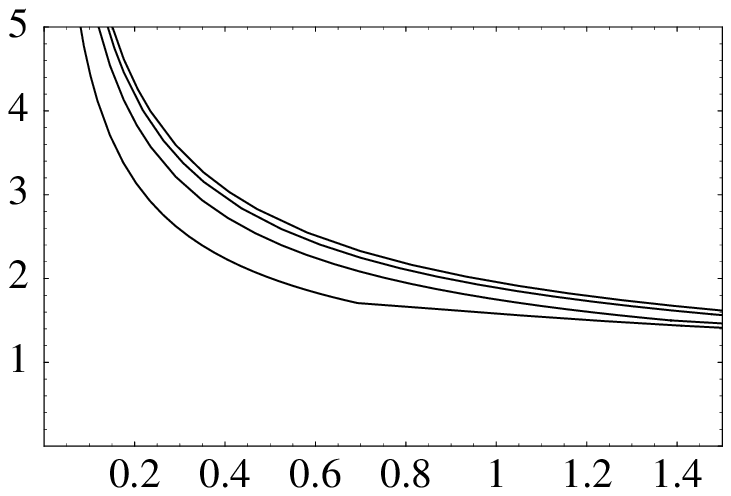}}
\put(1.8,3){\small{\darkgray $\Omega$}}
\put(6.2,0){\small{\darkgray $\tau_\tinyT$}}
\put(6.2,2.7){\small(a)}
\put(7.92,0.18){\includegraphics[width=5\unitlength]{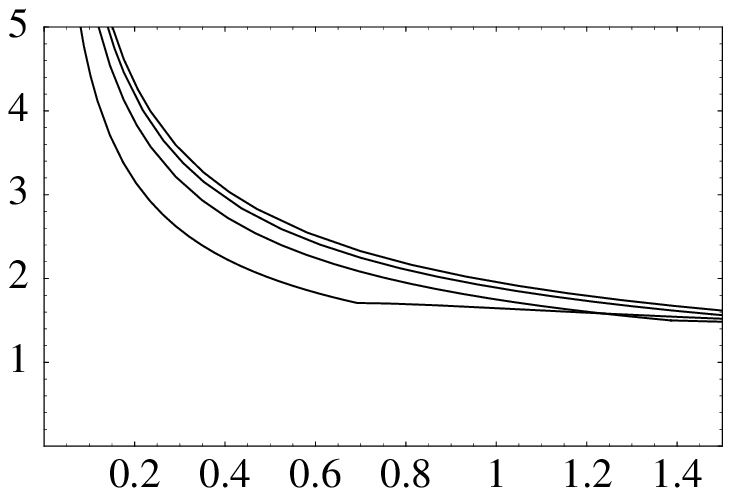}}
\put(7.8,3){\small{\darkgray $\Omega$}}
\put(12.2,0){\small{\darkgray $\tau_\tinyT$}}
\put(12.2,2.7){\small(b)}
\end{picture}
\end{center}
\caption{ 
  Plot of the thresholds $\Omega_{a,{\rm th}}$ in \refeq{th_a} and
  $\Omega_{c,{\rm th}}$ in \refeq{th_c} for different values of the
  number of clones. We fixed $\mu=0$ (a) [$\mu=0.4$ (b)] and, from
  bottom to top, we set $m=2,4,8,16$. The region above the lines refer
  to the case in which telecloning is more effective than the LCDT
  strategy.
\label{f:th}}
\end{figure}
In Fig.~\ref{f:th} the thresholds $\Omega_{a,{\rm th}}$ and
$\Omega_{c,{\rm th}}$ are plotted for different values of $m$ and
$\mu$ (see caption for details). The regions below the thresholds
refer to the regimes for which the local strategy is more effective
than telecloning. We notice that the benefits of telecloning become
slightly less effective as the number of modes $m$ and the thermal
noise $\mu$ increases. This is due to the fact that in this case the
performances of ideal telecloning decreases too.
\section{Conlusions}\label{s:esco}
In this paper we considered the application of CV cloning in a quantum
communication scenario. In particular, we analyzed an
amplitude-modulated channel in which a coherent signal has to be
distributed among $m$ parties. Based upon fidelity of the clones as a
figure of merit, we compared two strategies to perform the
distribution task in a noisy environment: on one hand the optimized
LCDT scheme, where no entanglement is present, on the other hand the
optimized telecloning protocol. Since the noise acts differently in
the two protocols, we found that telecloning is more effective than
the LCDT scheme for a wide range of noise parameters. This result
shows that entanglement, besides been recognized as a valuable
resource for a variety of two-party protocols, is a {\em robust}
resource also when many parties are involved.  Furthermore, the high
fidelity obtained by telecloning suggests that entanglement may be a
resource to enhance the exchanged information in a multiparty
communication network. Work along these lines is in progress and
results will be reported elsewhere.

\end{document}